\documentstyle[12pt]{article}
\title{Right-handed-neutrino Majorana mass at the SUSY GUT scale
and the solution of the solar-neutrino problem}
\author{L.\ Lavoura\thanks{On leave of absence from
Universidade T\'ecnica de Lisboa, Lisbon, Portugal} \\
\small Department of Physics, Carnegie-Mellon University, \\
\small Pittsburgh, Pennsylvania 15213, U.S.A.}
\begin{document}
\maketitle
\begin{abstract}
In the SUSY GUT scenario,
it is natural to assume the right-handed-neutrino Majorana-mass scale
to be $10^{16}$ GeV.
This will in principle lead,
by the seesaw mechanism,
to a $ \nu_{\tau} $ mass of order
$ m_t^2 / (10^{16}\, {\rm GeV}) \sim 3 \times 10^{-3}\, {\rm eV} $.
This suggests that the solution of the solar-neutrino puzzle
should be either the MSW effect in $\nu_e$--$\nu_{\tau}$ oscillations,
with $ m_{\nu_{\tau}}^2 \sim 10^{-5}\, {\rm eV}^2 $,
or long-wavelength $\nu_e$--$\nu_{\mu}$ oscillations,
with $ m_{\nu_{\mu}}^2 \sim 10^{-10}\, {\rm eV}^2 $.
These solutions require unexpectedly large mixings of $ \nu_e $
with $ \nu_{\tau} $ and $ \nu_{\mu} $,
respectively.
I suggest a variation of the Dimopoulos--Hall--Raby model
for the fermion mass matrices
which can accomodate such large mixings.
\end{abstract}

\vspace{2mm}

Supersymmetric Grand Unified Theories (SUSY GUT's)
have recently experienced an upsurge in popularity,
because they do not require the existence of an intermediate-energy
breaking of the grand-unification group
in order to obtain unification of the gauge couplings.
In a SUSY GUT,
the grand-unification gauge group is broken directly
to the standard-model gauge group at $ M_G \sim 10^{16}\, {\rm GeV} $,
while supersymmetry is broken at
a phenomenologically interesting scale $ M_{SUSY} \sim 10^3\, {\rm GeV} $
\cite{boer}.

The seesaw mechanism \cite{seesaw} for the suppression
of the left-handed-neutrino masses
can be incorporated in a SUSY GUT like SO(10).
That mechanism suggests that the mass of the $ \nu_{\tau} $
should be of order $ m_t^2 / M_I $,
where $ m_t $ is the top-quark mass
and $ M_I $ is the scale of the breaking
of the subgroup SU(4) \cite{patisalam} of SO(10).
This is because the right-handed-neutrino Majorana mass breaks SU(4),
and should therefore be of order $ M_I $;
while SU(4) relates the neutrino Dirac mass matrix
to the up-type-quark mass matrix,
and therefore the $ \nu_{\tau} $ Dirac mass should be of order $ m_t $.
The latter argument also suggests that the masses of the neutrinos
should have a strong hierarchy.

In the SUSY GUT scenario,
$ M_I $ should be equal to $ M_G $.
We therefore guess that $ m_{\nu_{\tau}} $
should turn out to be of order $ 3 \times 10^{-3}\, {\rm eV} $,
and that $ m_{\nu_e} \ll m_{\nu_{\mu}} \ll m_{\nu_{\tau}} $.
Such tiny neutrino masses would be uninteresting
from the points of view of direct laboratory measurements,
of a solution to the dark-matter problem,
and of a solution to the atmospheric-neutrino problem,
but that should be accepted as a price of the SUSY GUT scenario
\cite{ranfone}.

The solar-neutrino problem can be solved
by neutrino masses of these orders of magnitude in two different ways.
One may use a Mikheyev--Smirnov--Wolfenstein (MSW) \cite{msw}
resonant amplification of the $\nu_e$--$\nu_{\tau}$ oscillations
in the sun.
This requires \cite{mswfit}
$ m_{\nu_{\tau}}^2 \sim 10^{-5}\, {\rm eV}^2 $
and $ \sin^2 (2 \theta_{e \tau}) \sim 5 \times 10^{-3} $.
One may alternatively use long-wavelength vacuum
$\nu_e$--$\nu_{\mu}$ oscillations
\cite{justso}.
This requires \cite{justsofit}
$ m_{\nu_{\mu}}^2 \sim 10^{-10}\, {\rm eV}^2 $
and $ \sin^2 (2 \theta_{e \mu}) \sim 0.9 $ \cite{smirnov}.

Obtaining such large lepton-mixing angles is problematic
in the context of SO(10) models of the fermion mass matrices.
In those models,
the lepton mass matrices are related to the quark mass matrices.
One would then expect the lepton-mixing matrix $ K $
to have matrix elements of the same order of magnitude
as the corresponding matrix elements
of the quark-mixing matrix $ V $.
This would lead to $ \sin^2 (2 \theta_{e \mu}) \sim 10^{-1} $
and $ \sin^2 (2 \theta_{e \tau}) \sim 10^{-4} $,
mixing angles much too small for solving the solar-neutrino puzzle.

A nice model for the fermion mass matrices in a SUSY GUT
is the Dimopoulos--Hall--Raby (DHR) model \cite{DHR,DHR2}.
This model has great predictive power in the charged-fermion sector:
at the energy scale $ M_I = M_G $,
it predicts that $ m_b = m_{\tau} $,
$ m_s \approx m_{\mu} / 3 $,
$ m_d \approx 3 m_e $,
$ |V_{cb}| \approx \sqrt{m_c / m_t} $,
and $ |V_{ub} / V_{cb}| \approx \sqrt{m_u / m_c} $.
The top-quark mass must be quite high,
$ m_t \sim 180\, {\rm GeV} $,
both in order to get $ |V_{cb}| $ small enough,
and in order to obtain $ m_b = m_{\tau} $ at the scale $ M_I $,
by using the fixed-point structure
of the renormalization-group equations (RGE).

The DHR model has been extended to the neutrino sector \cite{DHRneutrinos}
and has encountered there the problem mentioned above.
Indeed,
in order to explain the solar-neutrino puzzle,
DHR have had to assume that it is the $\nu_e$--$\nu_{\mu}$ oscillations
which are MSW-enhanced.
That requires $ M_I \sim 10^{14}\, {\rm GeV} $,
two orders of magnitude smaller than $ M_G $.
This is inconsistent with the basic philosophy
of having a SUSY GUT without intermediate-energy symmetry breakings.

The purpose of this Brief Report is to suggest
a modification of the DHR scheme,
which keeps some of its predictive power
in the charged-fermion sector intact,
while solving the solar-neutrino problem with $ M_I = M_G $.
I have explored various modifications of the DHR scheme,
and have found that they all predict
too small a lepton mixing to be able to solve
the solar-neutrino problem in a consistent fashion.
The only exception that I have found is presented in this Brief Report.

I suggest that at $ M_I = M_G $ the fermion mass matrices are
\begin{eqnarray}
M_D & = &
\left( \begin{array}{ccc}
0 & a & d \\
a & b & 0 \\
d & 0 & c
\end{array} \right)\, ,
\label{eq:MD} \\
M_E & = &
\left( \begin{array}{ccc}
0 & a & - 3 d \\
a & - 3 b & 0 \\
- 3 d & 0 & c
\end{array} \right)\, ,
\label{eq:ME} \\
M_U & = &
\left( \begin{array}{ccc}
f & 0 & 0 \\
0 & 0 & s \\
0 & s & n
\end{array} \right)\, ,
\label{eq:MU} \\
M_{\nu N} & = &
\left( \begin{array}{ccc}
- 3 f & 0 & 0 \\
0 & 0 & s \\
0 & s & n
\end{array} \right)\, ,
\label{eq:MnuN} \\
M_{NN} & = &
\left( \begin{array}{ccc}
0 & 0 & d \\
0 & b & 0 \\
d & 0 & 0
\end{array} \right) \times r\, .
\label{eq:MNN}
\end{eqnarray}
These are,
respectively,
the mass matrix of the down-type quarks,
of the charged leptons,
of the up-type quarks,
the Dirac mass matrix of the neutrinos,
and the Majorana mass matrix of the right-handed neutrinos.
The effective Majorana mass matrix of the left-handed neutrinos is given,
in the seesaw approximation,
by
\begin{equation}
M_{\nu \nu} =
- M_{\nu N} M_{NN}^{-1} M_{\nu N}^T =
- \frac{1}{r}
\left( \begin{array}{ccc}
0 & -3 \frac{s f}{d} & - 3 \frac{n f}{d} \\
- 3 \frac{s f}{d} & 0 & 0 \\
- 3 \frac{n f}{d} & 0 & \frac{s^2}{b}
\end{array} \right)\, .
\label{eq:Mnunu}
\end{equation}
$ a $,
$ b $,
$ c $,
$ d $,
$ f $,
$ s $,
$ n $ and $ r $ are complex numbers.
$ a $ originates in the Yukawa couplings of the $ {\bf 10}^1 $ of SO(10).
$ c $ and $ n $ originate in the Yukawa couplings of the $ {\bf 10}^2 $,
a different $ {\bf 10} $.
$ s $ originates in the Yukawa couplings of the $ {\bf 10}^3 $.
$ b $ and $ d $ originate in the Yukawa couplings of the $ {\bf 126}^1 $.
$ f $ originates in the Yukawa couplings of the $ {\bf 126}^2 $.
The $ {\bf 126}^1 $ has a standard-model-breaking
vacuum expectation value (VEV) contributing to the mass matrices
of the down-type quarks and of the charged leptons,
and a standard-model-invariant VEV leading to $ M_{NN} $.
Similarly,
the $ {\bf 10}^2 $ has VEV's in two standard-model-breaking directions,
one VEV contributing to $ M_D $ and $ M_E $,
another one contributing to $ M_U $ and $ M_{\nu N} $.

The parameter $ r $ is the ratio between
the standard-model-invariant VEV of the $ {\bf 126}^1 $,
and its standard-model-breaking VEV which leads to a contribution
to $ M_D $ and to $ M_E $.
The latter VEV cannot be larger
than the effective VEV included in $ M_D $ and $ M_E $,
which is,
because below $ M_G $
we have the minimal supersymmetric standard model
with only two higgs doublets,
$ v \cos \beta = (175\, {\rm GeV}) \cos \beta $.
I now rely on the fact that the present model
makes predictions for the charged-fermion sector
which are very similar to the predictions of the DHR model,
and I rely on the DHR analysis \cite{DHR},
to conclude that $ \sin \beta $ should in the present model
be larger than $ 0.9 $,
a more likely value being much closer to $ 1 $ \cite{DHR2},
just as in the DHR model.
This means that $ v \cos \beta < 75\, {\rm GeV} $.
Therefore,
from the philosophy that SU(4) should be broken
at $ M_G \sim 10^{16}\, {\rm GeV} $,
I find $ |r| \sim 10^{15} $.

Notice that this model can be enforced by a simple $ Z_n $ symmetry,
$ n \geq 5 $,
on the Yukawa couplings.
With $ \omega^n = 1 $,
the $ {\bf 10}^1 $ transforms as $ \omega^{-3} $,
the $ {\bf 10}^2 $ transforms as $ \omega^{-6} $,
the $ {\bf 10}^3 $ transforms as $ \omega^{-5} $,
the $ {\bf 126}^1 $ transforms as $ \omega^{-4} $,
and the $ {\bf 126}^2 $ transforms as $ \omega^{-2} $.
The three lepton generations are in representations
$ {\bf \overline{16}}^{1,2,3} $
of SO(10);
the $ {\bf \overline{16}}^1 $ transforms as $ \omega^{1} $,
the $ {\bf \overline{16}}^2 $ transforms as $ \omega^{2} $,
and the $ {\bf \overline{16}}^3 $ transforms as $ \omega^{3} $.
I consider the existence of such a symmetry very important.
In general,
assuming the presence
of arbitrarily-located ``texture zeros'' in the mass matrices
is inconsistent from the field-theory point of view.
Often,
one may even prove that
there is certainly no symmetry capable of enforcing those zeros.

I now work out the form of the quark- and lepton-mixing matrices
in this model.
I perform the analysis at the energy $ M_G $,
at which energy the mass matrices are as in Eqs.~\ref{eq:MD} to \ref{eq:MNN}.
The purpose of the analysis is to show that
one gets at that energy scale values for the charged-fermion masses
and for the quark-mixing matrix
very similar to the ones in the DHR model.
If this is so,
the machinery of the RGE then runs those values down to the weak scale
in just the same way as it does in the DHR model,
and finally the agreement of both models with experiment
is just as good.
One does not need to run the mass matrix themselves,
as DHR have done \cite{DHR,DHR2};
it is equivalent,
but easier,
to perform the bi-diagonalizations
at the scale $ M_G $ and then to run the RGE for the masses
and for the mixing-matrix parameters down to the weak scale;
this procedure
allows one to work with a smaller set of differential equations
\cite{naculich}.

Let the orthogonal matrix $ R_U $ be such that
\begin{equation}
R_U^T
\left( \begin{array}{ccc}
|f| & 0 & 0 \\
0 & 0 & |s| \\
0 & |s| & |n|
\end{array} \right)
R_U = {\rm diag}(m_u, - m_c, m_t)\, .
\label{eq:MUdiagonal}
\end{equation}
Clearly,
$ |f| = m_u $,
$ |n| = m_t - m_c $,
and $ |s| = \sqrt{m_t m_c} $.
Therefore,
\begin{equation}
R_U =
\left( \begin{array}{ccc}
1 & 0 & 0 \\
0 & \sqrt{\frac{m_t}{m_t + m_c}} & \sqrt{\frac{m_c}{m_t + m_c}} \\
0 & - \sqrt{\frac{m_c}{m_t + m_c}} & \sqrt{\frac{m_t}{m_t + m_c}}
\end{array} \right)
\label{eq:RU}
\end{equation}
is a matrix which only mixes the second and third generations.
The unitary matrix $ U_D $ diagonalizes $ M_D M_D^{\dagger} $,
after some phases have been removed from that matrix:
\begin{equation}
U_D^{\dagger}
\left( \begin{array}{ccc}
|a|^2 + |d|^2 & |a b| & |c d| \\
|a b| & |a|^2 + |b|^2 & |a d| e^{i \psi} \\
|c d| & |a d| e^{- i \psi} & |c|^2 + |d|^2
\end{array} \right)
U_D = {\rm diag}(m_d^2, m_s^2, m_b^2)\, ,
\label{eq:MDdiagonal}
\end{equation}
where $ \psi \equiv \arg [ (a^2 c) / (b d^2) ] $.
Then,
the quark mixing matrix $ V $ is given by
\begin{equation}
V = R_U^T\, {\rm diag} (1, e^{i \theta}, 1)\, U_D\, ,
\label{eq:V}
\end{equation}
where $ \theta \equiv \arg [ (b d n) / (a c s) ] $.
I have used the fact that $ R_U $
only mixes the second and third generations
in order to eliminate one phase from $ V $,
by means of a rephasing of the up-quark field.

Similarly,
in the lepton sector,
let the unitary matrix $ R_{\nu} $ be such that
\begin{equation}
R_{\nu}^T
\left( \begin{array}{ccc}
0 & - 3 \left| \frac{s f}{d} \right| & - 3 \left| \frac{n f}{d} \right|
\\*[2mm]
- 3 \left| \frac{s f}{d} \right| & 0 & 0 \\*[2mm]
- 3 \left| \frac{n f}{d} \right| & 0 & \left| \frac{s^2}{b} \right|
\end{array} \right)
R_{\nu} = - r e^{i \arg (b / s^2)} {\rm diag}(m_1, m_2, m_3)\, .
\label{eq:Mnudiagonal}
\end{equation}
$ m_1 $,
$ m_2 $ and $ m_3 $ are real and positive,
with $ m_1 \leq m_2 \leq m_3 $,
and they are the light-neutrino masses.
The unitary matrix $ U_E $ diagonalizes $ M_E M_E^{\dagger} $,
after some phases have been removed from that matrix:
\begin{equation}
U_E^{\dagger}
\left( \begin{array}{ccc}
|a|^2 + 9 |d|^2 & - 3 |a b| & - 3 |c d| \\
- 3 |a b| & |a|^2 + 9 |b|^2 & - 3 |a d| e^{i \psi} \\
- 3 |c d| & - 3 |a d| e^{- i \psi} & |c|^2 + 9 |d|^2
\end{array} \right)
U_E = {\rm diag}(m_e^2, m_{\mu}^2, m_{\tau}^2)\, .
\label{eq:MEdiagonal}
\end{equation}
Then,
the lepton mixing matrix $ K $ is given by
\begin{equation}
K = R_{\nu}^T {\rm diag} (e^{i \xi} , e^{i \theta}, 1) U_E\, ,
\label{eq:K}
\end{equation}
where $ \xi \equiv \arg [ (d^2 s^2) / (b c f n) ] $.

We see that this model involves seven real numbers (excluding $ |r| $)
and three phases in the determination of the fermion masses
and of the mixing matrices,
as opposed to the DHR model,
which involves only six real numbers and one phase.

If $ d $ vanished,
$ M_D $ and $ M_E $ would be the same as in the DHR model.
$ U_D $ would only involve a first-second-generation mixing.
Because $ R_U $ only involves in the present model
a second-third-generation mixing (see Eq.~\ref{eq:RU}),
one concludes that a vanishing $ d $ would lead to a vanishing $ V_{ub} $,
which is excluded by experiment.
One may indeed show that in this model $ |V_{ub}| $
is proportional to $ |d| $:
\begin{eqnarray}
|V_{us}|^2 & \approx & \frac{m_d}{m_s + m_d} - \frac{|d|^2}{m_s m_b}
\left( \frac{m_d}{m_b} + \cos \psi \right)\, ,
\label{eq:Vus2}\\
|V_{ub}|^2 & \approx &
\frac{|d|^2}{m_b^2}\, .
\label{eq:Vub2}
\end{eqnarray}
Thus,
the Cabibbo angle agrees with experiment even if $ d=0 $,
but
\begin{equation}
|d|^2 \approx m_b^2 |V_{ub}|^2 \sim m_s m_d\, ,
\label{eq:destimate}
\end{equation}
which means that $ |d| $ and $ |a| $ are of the same order of magnitude.
{}From Eq.~\ref{eq:Vus2},
the result for the Cabibbo angle is not altered by the presence
of such a small $ d $,
and $ |V_{us}| \approx \sqrt{m_d / m_s} $.
$ U_D $ remains dominated by a first-second-generation mixing,
and therefore $ |V_{cb}| $ comes mostly from $ R_U $.
Just as in the DHR model,
$ |V_{cb}| \approx \sqrt{m_c / m_t} $ (see Eq.~\ref{eq:RU}),
and therefore the top-quark mass must be high.
The smallness of $ |d| $ also makes for it that
the relationships between the down-type-quark masses
and the charged-lepton masses are essentially unaltered
from what they were in the DHR model.
The main difference between the two models
is that $ |V_{ub} / V_{cb}| $ was predicted to be small
in the DHR model,
while in the present model it is largely arbitrary,
for it is proportional to the arbitrary parameter $ |d| $
(see Eq.~\ref{eq:Vub2}).

Now consider the mass of the heaviest neutrino.
As $ |s^2 / b| \gg |n f / d| \gg |s f / d| $,
we find
$ m_3 \approx (m_c m_t) / (m_s |r|) \sim 3 \times 10^{-3}\, {\rm eV} $
for $ |r| \sim 10^{15} $.
This is in the ball-park to explain the solar-neutrino deficit
via MSW-enhanced $\nu_e$--$\nu_{\tau}$ oscillations.
Notice however that this approximate formula is different
from the one that had been guessed in the beginning of this work.

Another interesting feature of the model is that
the mixing angle of the electron and tau neutrinos increases
when $ |V_{ub}| $ decreases,
while the naive analogy between the lepton- and the quark-mixing matrices
would have suggested the opposite.
Indeed,
$ |V_{ub}| $ is proportional to $ |d| $.
But,
when $ |d| $ decreases,
the $ (1, 3) $ and $ (2, 3) $ matrix elements of $ M_{\nu \nu} $
increase,
as seen in Eq.~\ref{eq:Mnunu};
this leads to larger mixings between $ \nu_e $ and $ \nu_{\tau} $,
and between $ \nu_{\mu} $ and $ \nu_{\tau} $
(but not between $ \nu_e $ and $ \nu_{\mu} $).

For the numerical work,
I have used the approximations suggested by Naculich \cite{naculich}.
The supersymmetry-breaking scale $ M_{SUSY} $
is fixed at $ 170\, {\rm GeV} $,
approximately equal to the top-quark mass.
The one-loop RGE of the supersymmetric standard model
are used to evolve the masses and mixing angles
from $ M_G $ down to $ M_{SUSY} $.
The gauge couplings are analytic functions of
$ t \equiv \ln (M / M_{SUSY}) / (16 \pi^2) $:
\begin{eqnarray}
g_1^2 (t) & = & \frac{40 \pi}{585 - 528 \pi t}\, ,
\nonumber\\
g_2^2 (t) & = & \frac{40 \pi}{301 - 80 \pi t}\, ,
\label{eq:gaugecouplings}\\
g_3^2 (t) & = & \frac{280 \pi}{687 + 1680 \pi t}\, .
\nonumber
\end{eqnarray}
They unify for $ t_G = 71 / (112 \pi) $,
at which scale they have the value $ g^2 (t_G) = 35 \pi / 219 $.
The unification scale is $ M_G = M_{SUSY} \exp (16 \pi^2 t_G) $.
For $ |r| $ I take,
in each particular case,
the value
\begin{equation}
|r| = \frac{M_G}{v \cos \beta}
= \frac{34}{35}\, \exp \left( \frac{71 \pi}{7} \right)\,
\sqrt{1 + \tan^2 \beta}\, .
\label{eq:rvalue}
\end{equation}
$ |r| $ is therefore a function of $ \beta $.
Notice that $ |r| $ might be larger than the value in Eq.~\ref{eq:rvalue},
for the standard-model-breaking VEV of the $ {\bf 126}^1 $ is unknown
and is only bounded to be smaller than $ v \cos \beta $.
$ |r| $ might therefore very well be one order of magnitude larger
than the value that I take for it,
which would correspondingly suppress all the light-neutrino masses.
At $ M_{SUSY} $,
the values of the quark- and charged-lepton masses are:
$ m_{\tau} = 1.749\, {\rm GeV} $,
$ m_{\mu} = 103.4\, {\rm MeV} $,
$ m_e = 500.9\, {\rm keV} $,
$ m_b = 2.89\, {\rm GeV} $,
$ m_s = 81\, {\rm MeV} $,
$ m_d = 4.1\, {\rm MeV} $,
$ m_c = 672\, {\rm MeV} $,
and $ m_u = 2.4\, {\rm MeV} $,
as obtained by Naculich from the QCD and QED running
of those masses from the energy scale at which they are known
up to $ M_{SUSY} $.
The quark-mixing parameters do not run significantly
at energy scales below $ M_{SUSY} $.
These are the values that I have tried to fit,
taking especial care to fit correctly
the charged-lepton masses and the Cabibbo angle.
The top-quark mass is $ m_t = 1.043\, m_t (M_{SUSY}) $,
due to the QCD correction.

Let us denote the squared eigenvalues of the Yukawa-coupling matrix
of the up-type quarks [$ M_U / (v \sin \beta) $] by $ U_{\alpha} $,
and the squared eigenvalues of the Yukawa-coupling matrices
of the down-type quarks and of the charged leptons
[$ M_D / (v \cos \beta) $ and $ M_E / (v \cos \beta) $,
respectively]
by $ D_i $ and $ E_i $,
respectively.
The indices $ \alpha $ and $ i $ take the values 1,
2 and 3.
These squared eigenvalues run with energy according to
\begin{eqnarray}
\frac{d U_{\alpha}}{d t} & = & U_{\alpha}
(a_U + b_M U_{\alpha} + 2 c_M \sum_i D_i |V_{\alpha i}|^2)\, ,
\nonumber\\
\frac{d D_i}{d t} & = & D_i
(a_D + b_M D_i + 2 c_M \sum_{\alpha} U_{\alpha} |V_{\alpha i}|^2)\, ,
\label{eq:eigenvaluerunning}\\
\frac{d E_i}{d t} & = & E_i (a_E + b_M E_i)\, .
\nonumber
\end{eqnarray}
The $ |V_{\alpha i}|^2 $ are the squared moduli of the quark-mixing-matrix
elements,
which also run with energy,
following \cite{sasaki}
\begin{eqnarray}
\frac{1}{c_M}\, \frac{d |V_{\alpha i}|^2}{d t} & = &
2 |V_{\alpha i}|^2 \left[
- U_{\alpha} - D_i
+ \sum_{\beta} U_{\beta} |V_{\beta i}|^2
+ \sum_j D_j |V_{\alpha j}|^2 \right.
\nonumber\\
                                             &   &
\left. + 2 U_{\alpha} D_i \left(
\sum_{\beta \neq \alpha} \frac{|V_{\beta i}|^2}{U_{\alpha} - U_{\beta}}
+ \sum_{j \neq i} \frac{|V_{\alpha j}|^2}{D_i - D_j} \right) \right]
\nonumber\\
                                             & + &
2 \sum_{\beta \neq \alpha} \sum_{j \neq i}
\left[ 2 {\rm Re} \left(
V_{\alpha i} V_{\beta j} V_{\alpha j}^{\ast} V_{\beta i}^{\ast}
\right) \right]\,
\left(
\frac{U_{\alpha} D_j}{U_{\alpha} - U_{\beta}}
+ \frac{U_{\beta} D_i}{D_i - D_j}
\right)\, .
\label{eq:CKMrunning}
\end{eqnarray}
Here \cite{sasaki,branco},
\begin{equation}
2 {\rm Re} \left(
V_{\alpha i} V_{\beta j} V_{\alpha j}^{\ast} V_{\beta i}^{\ast}
\right) =
1 - |V_{\alpha i}|^2 - |V_{\beta j}|^2 - |V_{\alpha j}|^2 - |V_{\beta i}|^2
+ |V_{\alpha i} V_{\beta j}|^2 + |V_{\alpha j} V_{\beta i}|^2
\label{eq:relation}
\end{equation}
for $ \beta \neq \alpha $ and $ j \neq i $.
The parameters $ a_U $,
$ a_D $,
$ a_E $,
$ b_M $ and $ c_M $ are model-dependent.
In the supersymmetric standard model,
$ c_M = 1 $ and $ b_M = 6 $,
and
\begin{eqnarray}
a_U & = & - \frac{26}{15} g_1^2 - 6 g_2^2 - \frac{32}{3} g_3^2
+ 6 (U_1 + U_2 + U_3)\, ,
\nonumber\\
a_D & = & - \frac{14}{15} g_1^2 - 6 g_2^2 - \frac{32}{3} g_3^2
+ 6 (D_1 + D_2 + D_3) + 2 (E_1 + E_2 + E_3)\, ,
\label{eq:auadae}\\
a_E & = & - \frac{18}{5} g_1^2 - 6 g_2^2
+ 6 (D_1 + D_2 + D_3) + 2 (E_1 + E_2 + E_3)\, .
\nonumber
\end{eqnarray}

I use as input the up-type-quark masses and the values of $ |a| $,
$ |b| $,
$ |c| $,
$ |d| $,
and the phases $ \psi $,
$ \theta $ and $ \xi $ at the energy $ M_G $.
Some typical fits are given in table 1.
\begin{table}
\centering
\begin{tabular}{||c|c|c|c|c|c|c|c|c|c|c|}
\hline
\hline
$ \tan \beta $ &
$ m_t $  &
$ m_c $  &
$ m_u $  &
$ |a| $  &
$ |b| $  &
$ |c| $  &
$ |d| $  &
$ \psi $  &
$ \theta $  &
$ \xi $  \\
\hline
\hline
$ 20 $ &
$ 225 $  &
$ 0.41 $ &
$ 0.0015 $ &
$ 0.00567 $ &
$ 0.0239 $ &
$ 1.25 $ &
$ 0.004 $ &
$ 30 $ &
$ 45 $ &
$ 180 $ \\
\hline
$ 7 $ &
$ 195 $  &
$ 0.39 $ &
$ 0.0014 $ &
$ 0.00553 $ &
$ 0.0232 $ &
$ 1.19 $ &
$ 0.009 $ &
$ 33 $ &
$ 45 $ &
$ 0 $ \\
\hline
$ 15 $ &
$ 200 $  &
$ 0.39 $ &
$ 0.0014 $ &
$ 0.00561 $ &
$ 0.0236 $ &
$ 1.22 $ &
$ 0.0054 $ &
$ 50 $ &
$ 40 $ &
$ 0 $ \\
\hline
\hline
\end{tabular}
\caption[]{Examples of sets of input values.
The phases $ \psi $,
$ \theta $ and $ \xi $ are in degrees;
all other values,
except $ \tan \beta $,
are in GeV.}
\end{table}

The neutrino masses and the lepton-mixing matrix
are calculated at $ M_G $ and do not run with energy.
For each of the fits given in Table 1,
the values of the neutrino masses and of the relevant lepton-mixing
parameters are given in Table 2.
\begin{table}
\centering
\begin{tabular}{||c|c|c|c|c|c|c|c|c|c|}
\hline
\hline
$ m_1 $ (eV) &
$ m_2 $ (eV) &
$ m_3 $ (eV) &
$ \sin^2 (2 \theta_{e \mu}) $  &
$ \sin^2 (2 \theta_{e \tau}) $  &
$ \sin^2 (2 \theta_{\mu \tau}) $  \\
\hline
\hline
$ 3.97 \times 10^{-6} $  &
$ 1.62 \times 10^{-5} $ &
$ 2.89 \times 10^{-3} $ &
$ 0.734 $ &
$ 4.86 \times 10^{-3} $ &
$ 7.49 \times 10^{-4} $ \\
\hline
$ 6.33 \times 10^{-6} $  &
$ 1.16 \times 10^{-5} $ &
$ 6.92 \times 10^{-3} $ &
$ 0.842 $ &
$ 3.29 \times 10^{-5} $ &
$ 4.42 \times 10^{-6} $ \\
\hline
$ 4.10 \times 10^{-6} $  &
$ 1.13 \times 10^{-5} $ &
$ 3.29 \times 10^{-3} $ &
$ 0.676 $ &
$ 1.08 \times 10^{-3} $ &
$ 1.47 \times 10^{-4} $ \\
\hline
\hline
\end{tabular}
\caption[]{Neutrino-masses and mixing parameters yielded
by the three examples in Table 1,
respectively.}
\end{table}
For the calculation of the neutrino masses,
I have assumed $ |r| $ to be given by Eq.~\ref{eq:rvalue}.
Also,
$ \sin^2 (2 \theta_{e \mu}) \equiv
4 |K_{1 e} K_{2 \mu} K_{1 \mu} K_{2 e}| $,
with similar definitions for $ \sin^2 (2 \theta_{e \tau}) $
and for $ \sin^2 (2 \theta_{\mu \tau}) $.

One sees that it is possible to obtain $ \theta_{e \tau} $
sufficiently large to allow for the solar-neutrino problem
to be solved by MSW-enhanced $\nu_e$--$\nu_{\tau}$ oscillations.
Also,
$ \sin^2 (2 \theta_{e \mu}) $ is quite large,
and
it may be as large
as to allow a solution of the solar-neutrino puzzle
by means of $\nu_e$--$\nu_{\mu}$ vacuum oscillations.
The neutrino masses are in general adequate for those explanations
of the solar-neutrino deficit.
One sees that $ m_2 / m_1 \sim 3 $ and $ m_3 / m_2 \sim 300 $.

In the present model,
a solution of the solar-neutrino problem purely by
$\nu_e$--$\nu_{\mu}$ vacuum oscillations is possible.
The mixing angle $ \theta_{e \tau} $ can be made sufficiently small
to render the MSW effect between $ \nu_e $ and $ \nu_{\tau} $
irrelevant.
This is what happens
in the second example in Tables 1 and 2.
On the other hand,
a solution of the solar-neutrino problem solely by the MSW effect
between $ \nu_e $ and $ \nu_{\tau} $,
though possible if one just takes into account the masses
and the mixing angle of those two neutrinos
(see the first example in Tables 1 and 2),
should not be attempted.
This is because the mass difference $ m_2^2 - m_1^2 $
is in general in the right range
to suppress the $^8{\rm B}$ neutrinos
by means of long-wavelength vacuum oscillations,
and the mixing angle $ \theta_{e \mu} $ is always so large
that the effect of those vacuum oscillations cannot be neglected.
It is possible to find examples
(with high $ \tan \beta $)
in which $ m_2^2 - m_1^2 $ is so small
as not to suppress the $ ^8{\rm B} $ neutrino signal meaningfully
by means of the vacuum oscillations
(the $ pp $ neutrino signal will still be suppressed);
but in those examples,
because $ m_3 / m_2 \sim 300 $ in this model,
$ m_3^2 - m_1^2 $ will also be too small for the MSW effect
to be able to explain the solar-neutrino deficit.

In general,
a mixed solution of the solar-neutrino problem
should be considered.
An example which requires such a mixed interpretation
of the solar-neutrino deficit is the third one in Tables 1 and 2.
There,
$ m_3^2 - m_1^2 \approx 10^{-5}\, {\rm eV}^2 $ and
$ \sin^2 (2 \theta_{e \tau}) \approx 10^{-3} $
are such that the MSW effect provides
a suppression of the solar-neutrino signal
in all present experiments (Homestake,
Kamiokande and gallium) equal to about 50\% of the suppression
which is actually observed.
The other half of the observed suppression
is attributed to the $\nu_e$--$\nu_{\mu}$ vacuum oscillations,
with $ m_2^2 - m_1^2 \approx 1.1 \times 10^{-10}\, {\rm eV}^2 $
and $ \sin^2 (2 \theta_{e \mu}) \approx 0.68 $.
In general,
in the present model one should consider both the
MSW effect in the $\nu_e$--$\nu_{\tau}$ oscillations,
and the long-wavelength $\nu_e$--$\nu_{\mu}$ oscillations,
when attempting to explain the solar-neutrino deficit,
because both effects will suppress the solar-neutrino signal
significantly.

In conclusion,
I have argued in this Brief Report that in the SUSY GUT scenario
the neutrino masses should be extremely small,
in such a way that the MSW explanation of the solar-neutrino deficit
cannot be through $\nu_e$--$\nu_{\mu}$ oscillations,
but can quite likely occur through $\nu_e$--$\nu_{\tau}$ oscillations.
Vacuum $\nu_e$--$\nu_{\mu}$ oscillations
are also a possible explanation of the solar-neutrino deficit
in this scenario.
As an illustration,
I have constructed a mass-matrix model
which yields predictions for the charged-fermion sector
similar to the ones of the DHR model,
but which is able to fit the relatively large mixing angles
needed in these alternative views of the solar-neutrino suppression.

\vspace{5mm}

I thank Lincoln Wolfenstein for first calling my attention
to the problem of the overall scale of the neutrino masses
in SUSY GUT models in general,
and in the DHR model in particular.
I also thank him for discussions,
and for reading the manuscript.
This work was supported by the United States Department of Energy,
under the contract DE-FG02-91ER-40682.


%
%
\end{document}